\documentclass{article}

\usepackage{arxiv}

\usepackage[utf8]{inputenc} 
\usepackage[T1]{fontenc}    
\usepackage{hyperref}       
\usepackage{url}            
\usepackage{booktabs}       
\usepackage{amsfonts}       
\usepackage{nicefrac}       
\usepackage{microtype}      
\usepackage{graphicx}
\usepackage[numbers,sort&compress]{natbib}
\usepackage{doi}
\usepackage{amsmath}
\usepackage{amssymb}
\usepackage{algorithm}
\usepackage{algorithmic}
\usepackage{textcomp}
\usepackage{booktabs}
\usepackage{array}
\usepackage{enumitem}
\usepackage{float} 
\usepackage{placeins}
\usepackage{cleveref}       

\title{Quantum Computing for RNA Secondary Structure Prediction: An Adaptive Layered CVaR-VQE Framework with Comprehensive Evaluation of Optimization Algorithms}

\author{
    Sahebeh Nabavi \\
    Computer Engineering Department \\
    Sharif University of Technology \\
    Tehran, Iran \\
    \texttt{sahebeh.nabavi23@sharif.edu}
    \And
    Somayyeh Koohi \\
    Computer Engineering Department \\
    Sharif University of Technology \\
    Tehran, Iran \\
    \texttt{koohi@sharif.edu}
}

\renewcommand{\shorttitle}{Adaptive Layered CVaR-VQE for RNA Secondary Structure Prediction}

\hypersetup{
pdftitle={Quantum Computing for RNA Secondary Structure Prediction},
pdfsubject={quant-ph, cs.ET, q-bio.BM},
pdfauthor={Seyedeh Sahebeh Nabavi, Somayyeh Koohi},
pdfkeywords={Quantum Computing, Layered CVaR-VQE, RNA Secondary Structure Prediction},
}

\begin{document}
\maketitle

\begin{abstract}
Predicting RNA secondary structure remains a fundamental challenge in computational biology, as classical minimum free energy approaches face NP-hard complexity. While quantum computing offers promising alternatives, existing quantum methods predominantly rely on fixed variational quantum eigensolver architectures with limited adaptability to sequence-specific structural complexity. This paper introduces an adaptive gate-based quantum framework for RNA secondary structure prediction that dynamically adjusts circuit depth through a layered Conditional Value-at-Risk VQE architecture. Unlike conventional approaches, our method designs problem-specific ansatz circuits that encode RNA base-pair interactions into qubit states and iteratively refines energy estimations through a hybrid quantum-classical optimization loop. We conduct a comprehensive benchmarking of seven classical and nature-inspired optimization algorithms, namely Genetic Algorithm, Particle Swarm Optimization, Pattern Search, Simulated Annealing, Surrogate Optimization, Simultaneous Perturbation Stochastic Approximation, and COBYLA, on a benchmark dataset of RNA sequences. The evaluation is performed against three standard baseline methods, namely ViennaRNA, MATLAB MFE, and ProbKnot. Extensive simulations reveal that population-based and direct-search methods significantly outperform gradient-dependent and surrogate-based approaches. Specifically, Genetic Algorithm, Particle Swarm Optimization, and Pattern Search achieve the highest mean sensitivity of 70\%, 70\%, and 67\%, and mean specificity of 66\%, 70\%, and 61\% against ViennaRNA, respectively. Layer-wise analysis demonstrates that intermediate circuit depths consistently yield superior accuracy compared to the first layer, indicating that adaptive layering effectively captures structural complexity without the noise susceptibility of overly deep circuits. This work represents the first comprehensive comparative study of optimization algorithms for quantum RNA secondary structure prediction and the first application of an adaptive layered quantum approach to this problem, providing critical insights into algorithm-circuit co-design and establishing a scalable foundation for future hybrid quantum-classical bioinformatics pipelines.
\end{abstract}


\section{Introduction}
\label{sec:introduction}

The intricate three-dimensional architecture of Ribonucleic acid (RNA) dictates its biological function. Beyond its canonical role as an intermediary between DNA and proteins, RNA participates in gene regulation, catalysis, and cellular signaling. The secondary structure of RNA, which arises from intramolecular Watson-Crick and wobble base pairings, serves as a critical intermediate between the linear sequence and the functional 3D conformation. Understanding this secondary structure is essential for deciphering RNA function and accelerating therapeutic development, particularly in the context of mRNA vaccines and RNA interference. Classical computational tools, such as Mfold and RNAfold, rely on thermodynamic models to predict the minimum free energy (MFE) conformation \citep{higgs2000rna, mathews2006prediction, mathews2006revolutions, reeder2006beyond, seetin2012rna}. However, determining the MFE structure is an NP-hard problem \citep{bonnet2020designing, aguirre2007computational}. Dynamic programming algorithms, including those developed by Nussinov and Zuker \citep{nussinov1980fast, zuker1981optimal}, suffer from exponential growth in computational complexity as sequence length increases, severely limiting their applicability to large, biologically relevant RNA molecules with complex pseudoknots.

The structural flexibility and functional diversity of RNA molecules are deeply intertwined. RNA can adopt a wide array of conformations, from simple hairpin loops to intricate pseudoknotted structures. These secondary structural motifs are not merely static scaffolds; they are dynamic elements that govern interactions with proteins, other nucleic acids, and small molecules \citep{trigiante2021emerging, cao2024identification, zhang2022advances}. Misfolding or disruptions in these structural elements can lead to loss of function and are implicated in various diseases. Therefore, accurate prediction of RNA secondary structure is not just a computational exercise but a fundamental step toward understanding RNA biology at a molecular level \citep{vicens2022thoughts, low2010shape}. While classical thermodynamics-based models have provided invaluable insights, they often fall short when dealing with long sequences, as the search space grows exponentially, making exact optimization computationally prohibitive \citep{ray2012rna}. Machine learning and deep learning models have recently been explored as alternatives \citep{zhao2021review, sato2023recent, saman2022rna, chen2023redfold, wu2023machine, zhang2024predicting, upadhyay2026deep, sacco2026machine}, in the same spirit as data-driven structure prediction for proteins \citep{jumper2021highly}, although they depend on large curated training sets.

The advent of quantum computing offers a promising paradigm shift for addressing such combinatorial optimization problems \citep{pal2024quantum}. Quantum algorithms leverage superposition and entanglement to explore vast solution spaces simultaneously, providing a theoretical advantage over classical sequential exploration. In the context of RNA folding, quantum approaches can navigate the complex, non-convex energy landscape more effectively. Early explorations in this domain have primarily focused on two hardware paradigms: quantum annealing and gate-based quantum computing. Quantum annealers, such as those developed by D-Wave, are special-purpose devices. They evolve a physical system from a simple initial Hamiltonian toward a problem Hamiltonian and aim to find its lowest-energy state. This makes them suitable for optimization, and specific QUBO formulations of RNA folding have been mapped onto them \citep{fox2022rna, zaborniak2022qubo}. Their scope, however, is limited: they cannot be programmed to run arbitrary algorithms.

\begin{figure}[!htbp]
    \centering
    \includegraphics[width=0.8\columnwidth]{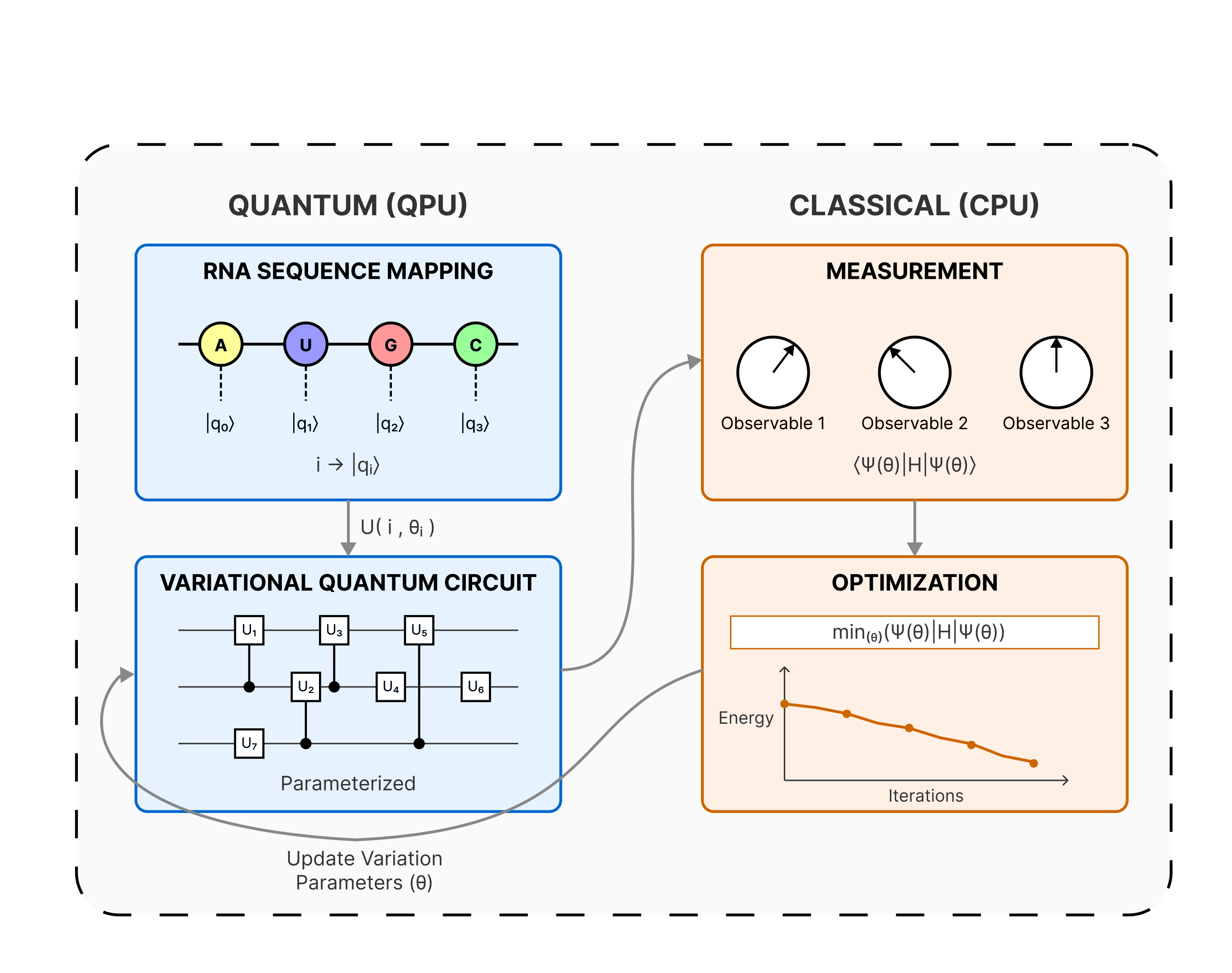}\vspace{-6pt}
    \caption{Overview of the hybrid VQE workflow for RNA folding. On the quantum processing unit (QPU), the nucleotide sequence is mapped to qubits and a parameterized circuit \(U(i,\theta_i)\) prepares the trial state \(|\Psi(\theta)\rangle\). On the classical processor (CPU), measured expectation values of the Hamiltonian \(H\) are used to update \(\theta\) through a classical optimizer, and the loop repeats until the energy stops decreasing.}
    \label{fig:vqe_flow}
\end{figure}

Gate-based quantum computers, such as those from IBM, Google, and IonQ, are general-purpose machines. They are built from elementary quantum gates applied to qubits in a circuit, which makes them universal in principle and applicable to many problem types, not only optimization. Their gate-level control also allows the same hardware to run hybrid variational algorithms such as the Variational Quantum Eigensolver (VQE) and the Quantum Approximate Optimization Algorithm (QAOA) \citep{jiang2023predicting, khatami2023gate, casares2022qfold, robert2021resource, chandarana2023digitized, luo2020quantum, doga2024perspective}. The price of this flexibility is circuit depth and noise, which limit the problem sizes that can be run today. This trade-off motivates the gate-based approach taken in this work. Compared with quantum annealing, which is tied to optimization problems, gate-based quantum computers are general-purpose: they can in principle run any algorithm that can be written as a quantum circuit, not only optimization tasks. This general-purpose nature is the main reason for choosing a gate-based approach in this work.

This work adopts the Variational Quantum Eigensolver (VQE), a hybrid algorithm that runs on a gate-based quantum computer, to address this problem. Variational quantum algorithms iteratively refine a parameterized quantum state to minimize an objective function—in this case, the RNA free energy—through a classical-quantum feedback loop. Figure~\ref{fig:vqe_flow} shows this loop schematically: the quantum processor prepares the trial state, while the classical processor measures the energy and updates the parameters. The Conditional Value-at-Risk (CVaR-VQE) variant further improves sampling efficiency by focusing on the lowest-energy tail of the probability distribution \citep{zaborniak2022qubo, zhang2024resource}. Related quantum approaches have also been applied to mRNA secondary structure and codon optimization \citep{kumar2025mrna, alevras2025codon, fox2021mrna}, RNA 3D motif identification \citep{fang2025rna}, and DNA sequence reconstruction \citep{huh2024quantum, sarkar2021quaser}. Despite these advancements, directly optimizing the RNA energy landscape remains challenging due to the high dimensionality, non-convexity, and susceptibility to noise. Traditional classical optimizers, such as genetic algorithms, differential evolution, and the Nakanishi-Fujii-Todo (NFT) method, often struggle with slow convergence and local minima \citep{chen2000prediction, montaseri2014rna, nakanishi2020sequential}. To the best of our knowledge, a comprehensive comparative study of classical optimization algorithms within a gate-based CVaR-VQE framework for RNA folding has not been conducted. Furthermore, existing approaches predominantly rely on fixed-depth ansatz circuits, which lack the adaptability required to capture the varying structural complexity of different RNA sequences. A summary of the prominent quantum RNA folding studies and their respective capabilities is presented in Table~\ref{tab:comparison}.

\begin{table*}[!htbp]
\centering
\caption{Comparison of Quantum Approaches for RNA Secondary Structure Prediction}
\label{tab:comparison}
\vspace{8pt}
\renewcommand{\arraystretch}{1.4} 
\resizebox{\textwidth}{!}{%
\begin{tabular}{|c|c|c|c|c|p{5.5cm}|}
\hline
\textbf{Study} & \textbf{Quantum Computer Model} & \textbf{Optimization Algorithm} & \textbf{RNA Length} & \textbf{Qubits (Q)} & \textbf{Main Contribution} \\
\hline
Fox \textit{et al.} \citep{fox2022rna} & Quantum Annealing & D-Wave Custom & $\sim$45 & 45 & First RNA folding on a quantum annealer; includes pseudoknots. \\
\hline
Zaborniak \textit{et al.} \citep{zaborniak2022qubo} & Quantum Annealing & SPSA & 20--50 & 20--50 & QUBO formulation for variational hybrid annealing. \\
\hline
Jiang \textit{et al.} \citep{jiang2023predicting} & Gate-Based & QAOA & 22--27 (Sim) & 22--27 & First gate-based RNA folding study. \\
\hline
Alevras \textit{et al.} \citep{alevras2024mrna} & Gate-Based & NFT & 15--60 & 10--80 & Utility-scale quantum computer (up to 42 nt) with CVaR-VQE. \\
\hline
\textbf{This Work} & Gate-Based & \textbf{7 Algorithms*} & 9--25 & 6--18 & \textbf{First adaptive layered approach}; comprehensive optimizer comparison in Layer-CVaR-VQE. \\
\hline
\end{tabular}}
\vspace{2mm}
\footnotesize{\\ *Algorithms compared: GA, PSO, PS, SPSA, SA, Surrogate, and COBYLA.}
\end{table*}

To address these gaps, this paper introduces a novel, adaptive gate-based framework for RNA secondary structure prediction. Our approach builds directly on Layer VQE \citep{liu2022layer}, which was originally proposed for combinatorial optimization on noisy quantum devices and grows the circuit one layer at a time; we adapt this strategy to the RNA folding Hamiltonian and combine it with a CVaR objective \citep{barkoutsos2019improving}. Our primary contribution is the implementation of a \textbf{Layer-CVaR-VQE} architecture, which dynamically adjusts the circuit depth through a layered ansatz. This represents the first application of an adaptive layered quantum approach to the RNA secondary structure prediction problem. Additionally, we conduct an extensive benchmarking of seven classical and nature-inspired optimization algorithms—Genetic Algorithm (GA), Particle Swarm Optimization (PSO), Pattern Search (PS), Simulated Annealing (SA), Surrogate Optimization (SUR), Simultaneous Perturbation Stochastic Approximation (SPSA), and COBYLA—within the variational pipeline on a benchmark set of RNA sequences. The performance is evaluated against three standard baseline methods: ViennaRNA \citep{ViennaRNA}, MATLAB MFE \citep{MATLABVQE}, and ProbKnot \citep{chakraborty2023quantum}. 

The behavior of the optimizer within the CVaR-VQE loop is critical, as the objective function is highly non-linear and involves discrete bitstring evaluations subject to shot noise. Gradient-based methods often fail in such landscapes, whereas population-based methods can explore the parameter space more robustly. Our empirical results demonstrate that the Layer-CVaR-VQE architecture yields superior sensitivity and specificity compared to the standard single-layer CVaR-VQE for the vast majority of optimizers, particularly at intermediate circuit depths (Layer 2 or Layer 3). This validates the hypothesis that a certain level of ansatz expressivity is required to capture structural correlations, but excessive depth introduces noise. Among the optimizers, GA, PSO, and PS achieve the highest mean sensitivities of 70.0\%, 70.0\%, and 67.0\%, and mean specificities of 66.0\%, 70.0\%, and 61.0\% against ViennaRNA, respectively. The superiority of these methods is attributed to their derivative-free global exploration capabilities. Furthermore, the quantum state probability analysis reveals that GA and PSO achieve higher mean probabilities (up to 32.0\% and 25.0\%), indicating a robust amplitude concentration on the optimal bitstrings.

The main contributions of this work are summarized as follows:
\begin{itemize}[leftmargin=*, nosep]
    \item \textbf{Adaptive Layered Ansatz:} Building on the Layer VQE strategy of \citet{liu2022layer}, we propose the first adaptive layered quantum approach for RNA secondary structure prediction, demonstrating that intermediate circuit depths (Layer 2 and 3) consistently outperform the standard fixed-depth (Layer 1) CVaR-VQE.
    \item \textbf{Comprehensive Optimizer Benchmarking:} We provide the first extensive comparative study of seven optimization algorithms within the Layer-CVaR-VQE pipeline, identifying GA, PSO, and PS as the most effective optimizers for this specific quantum biological problem.
    \item \textbf{Quantum Probability Insights:} We analyze the quantum state probability distributions, showing that population-based methods (GA and PSO) achieve superior amplitude concentration on the most stable bitstrings, whereas gradient-based methods like COBYLA fail to navigate the stochastic landscape.
    \item \textbf{Baseline Comparison:} We rigorously evaluate the proposed framework against three widely adopted classical baselines (ViennaRNA, MATLAB MFE, and ProbKnot) across a set of test sequences, establishing the competitiveness of the hybrid quantum approach.
\end{itemize}

The remainder of this paper is organized as follows. Section~\ref{sec:method} details the proposed Layer-CVaR-VQE methodology and the quantum circuit design. Section~\ref{sec:results} presents the experimental results and a comprehensive discussion. Finally, Section~\ref{sec:conclusion} concludes the paper and outlines future directions.

\section{Method}
\label{sec:method}

This section details the mathematical formulation of the RNA secondary structure prediction problem and the proposed quantum-classical framework. We begin by explaining the mapping of the RNA sequence to a quantum Hamiltonian and the fundamental principles of the Variational Quantum Eigensolver (VQE). Subsequently, we justify the choice of the gate-based quantum model and introduce the adaptive Layer-CVaR-VQE architecture. Finally, we discuss the integration of the Conditional Value-at-Risk (CVaR) objective and the rationale behind benchmarking multiple classical optimizers to navigate the complex energy landscape.

\subsection{Background and Quantum Formulation}
The core objective of the Variational Quantum Eigensolver (VQE) is to find the ground-state energy of a given physical system. In quantum mechanics, the state of a system is described by the Schr\"odinger equation:
\begin{equation}
    H |\psi\rangle = E |\psi\rangle
\end{equation}
where \(H\) is the Hamiltonian operator of the system, \(|\psi\rangle\) is the quantum state, and \(E\) is the associated energy eigenvalue. To find the minimum energy (ground state), VQE relies on the variational principle. By preparing a parameterized trial state \(|\psi(\theta)\rangle\) using a quantum circuit, the expectation value of the Hamiltonian, \(\langle \psi(\theta) | H | \psi(\theta) \rangle\), provides an upper bound to the true ground-state energy. Writing \(H=\sum_k E_k |E_k\rangle\langle E_k|\) with \(E_0 \le E_1 \le \dots\), this bound holds for every normalized trial state, with equality only when the trial state lies entirely in the ground-state subspace. A classical optimizer iteratively updates the parameters \(\theta\) to minimize this expectation value.

\begin{figure*}[!htbp]
    \centering
    \includegraphics[width=\textwidth, height=0.9\textwidth, keepaspectratio]{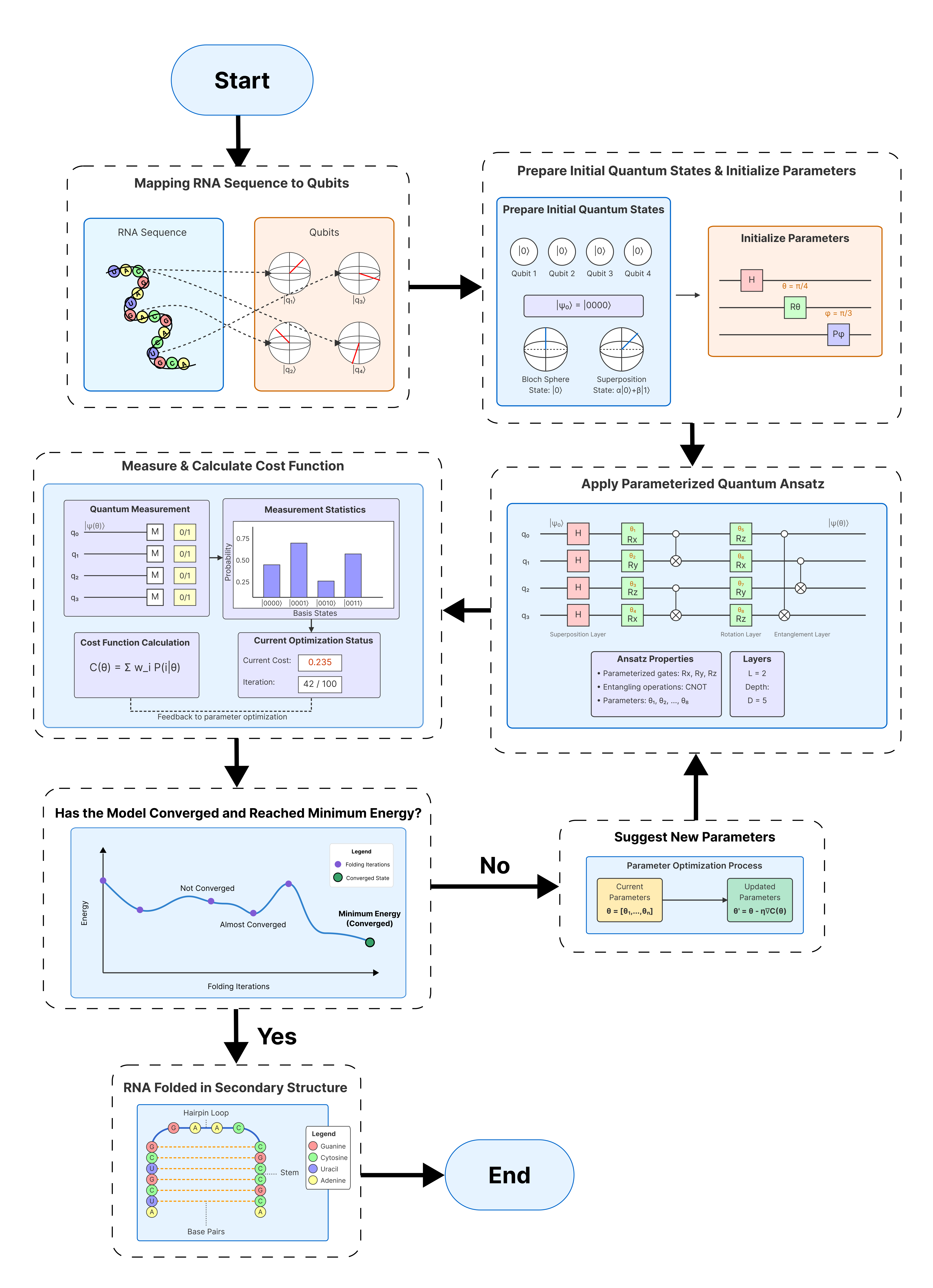}
    \caption{The step-by-step operational workflow of the adaptive Layer-CVaR-VQE for RNA secondary structure prediction. The diagram illustrates the transition from classical sequence mapping to quantum state preparation, parameterized ansatz application, and the classical optimization feedback loop.}
    \label{fig:process}
\end{figure*}

In the context of RNA secondary structure prediction, the problem is mapped to a Quadratic Unconstrained Binary Optimization (QUBO) formulation. The RNA sequence is first analyzed to identify all valid Watson-Crick and wobble base pairs.Watson-Crick pairs are the canonical hydrogen-bonded pairs: C-G (three hydrogen bonds) and A-U (two hydrogen bonds). Wobble pairs, mainly G-U, are weaker but still allowed in RNA. Let \(Q\) be the total number of valid candidate pairs. We assign one qubit to each candidate pair. The state of the \(q\)-th qubit, \(Z_q \in \{0, 1\}\), represents whether the \(q\)-th base pair is formed (1) or broken (0). The Hamiltonian of the RNA folding problem, \(H_{\text{RNA}}\), is constructed to reward valid structural motifs and penalize unphysical conformations. To maintain readability, the Hamiltonian is expressed as follows:
\begin{equation}
\begin{split}
    H_{\text{RNA}} = & \sum_{q=1}^{Q} h_q Z_q + \sum_{q, r \in S} J_{qr}^{\text{stack}} Z_q Z_r \\
    & + \sum_{q, r \in R} J_{qr}^{\text{reuse}} Z_q Z_r + \sum_{q, r \in C} J_{qr}^{\text{cross}} Z_q Z_r
\end{split}
\end{equation}
Here, \(Z_q\) is the Pauli-Z operator acting on the \(q\)-th qubit. The four terms in the Hamiltonian represent the following physical and structural constraints:
\begin{itemize}[leftmargin=*, nosep]
    \item \textbf{Base Pair Penalty (\(h_q\)):} Assigned to each valid pair to represent its thermodynamic stability. 
    \item \textbf{Stacking Reward (\(S\)):} A negative coefficient (\(J_{qr}^{\text{stack}} < 0\)) is applied when two consecutive pairs (e.g., \((i,j)\) and \((i+1, j-1)\)) form a stable stem.
    \item \textbf{Reuse Penalty (\(R\)):} A positive coefficient (\(J_{qr}^{\text{reuse}} > 0\)) is added to prevent a nucleotide from participating in more than one base pair simultaneously (steric clash).
    \item \textbf{Crossing Penalty (\(C\)):} A positive coefficient (\(J_{qr}^{\text{cross}} > 0\)) is applied to penalize crossing pairs, thereby suppressing pseudoknots and enforcing planar secondary structures.
\end{itemize}

By translating these rules into the Hamiltonian, the RNA folding problem becomes equivalent to finding the ground state of \(H_{\text{RNA}}\). The general workflow of this hybrid quantum-classical process is illustrated in Figure~\ref{fig:process}. 

As depicted in Figure~\ref{fig:process}, the operational flow begins with mapping the RNA sequence to a specific set of qubits based on valid base pairs. The initial states are prepared, and the parameterized ansatz is applied to explore the Hilbert space. The measurement statistics are then fed into the CVaR cost function, which guides the classical optimizer to update the parameters. This hybrid loop continues until the minimum energy is found, yielding the final secondary structure.

\subsection{Quantum Encoding and State Decoding}
To evaluate the RNA folding problem on a quantum computer, the biological sequence must first be translated into a quantum-mechanical representation. The encoding process begins by scanning the RNA sequence to identify all valid Watson-Crick and wobble base pairs (A-U, U-A, C-G, G-C, and G-U). A minimum loop length constraint is also applied, ensuring that a pair can only form if the distance between the two nucleotides is strictly greater than two (\(j - i > 2\)). This prevents sterically impossible conformations.

Let \(Q\) be the total number of valid candidate pairs found in the sequence. We map each candidate pair to a single qubit. Consequently, the quantum circuit requires exactly \(Q\) qubits. A computational basis state (a bitstring) of length \(Q\) directly represents a potential secondary structure: a state of \(|1\rangle\) (or bit '1') indicates that the corresponding base pair is formed, while \(|0\rangle\) (bit '0') indicates that it remains unpaired. 

To illustrate this mapping, consider a short RNA sequence \(5'\)-C A G U C G-3' with a length of \(N=6\). The algorithm scans the sequence and identifies three valid candidate pairs that satisfy the distance constraint:
\begin{itemize}[leftmargin=*, nosep]
    \item \(q_1\): Pair (1, 6) \(\rightarrow\) C-G
    \item \(q_2\): Pair (2, 5) \(\rightarrow\) A-U
    \item \(q_3\): Pair (3, 4) \(\rightarrow\) G-C
\end{itemize}
Since there are \(Q=3\) valid pairs, the problem is mapped onto a 3-qubit system. A bitstring of \(|111\rangle\) represents the case where all three pairs are formed simultaneously, creating a perfect stem. Conversely, \(|000\rangle\) represents a completely unfolded sequence.

After the quantum circuit is executed and measured, the classical optimizer receives a probability distribution over all possible bitstrings. The most probable bitstring (or the one with the lowest energy according to the Hamiltonian) is selected as the final structural prediction. This bitstring is then decoded back into the dot-bracket notation. For instance, if the measurement yields the bitstring \(|101\rangle\), it means \(q_1\) and \(q_3\) are active. According to the mapping, the pair (1, 6) forms an outer stem, but the pair (2, 5) is broken, leaving the nucleotides at positions 2 and 5 unpaired. The nucleotide at position 3 pairs with position 4, creating a smaller nested loop. The resulting dot-bracket structure is `( . ( ) . )`. This decoding step bridges the gap between the abstract quantum state and the biologically meaningful secondary structure.

\subsubsection{Ansatz Circuit Construction}
The expressivity of the VQE heavily relies on the design of the parameterized quantum circuit, known as the ansatz. In this work, we utilize a hardware-efficient layered ansatz, following the layer-by-layer construction of Layer VQE \citep{liu2022layer}. The construction of the circuit begins with a superposition layer, where a parameterized \(R_y(\theta)\) rotation gate is applied to each qubit. This allows the qubits to explore multiple states simultaneously. 

Following the rotation layer, we apply an entangling block consisting of a series of controlled-NOT (CNOT) or controlled-Z (CZ) gates arranged in a linear topology. This entangling layer creates correlations between neighboring qubits, which is essential for representing the stacking interactions and loop closures in the RNA structure. For a Layer-VQE architecture with \(L\) layers, the sequence of \(R_y\) rotations and CNOT entanglements is repeated \(L\) times. The total number of trainable parameters in the circuit is exactly \(L \times Q\). As we increment the number of layers, the circuit gains the ability to represent more complex, long-range correlations, though at the cost of a deeper circuit susceptible to noise.

\subsection{Choice of Quantum Hardware: Gate-Based vs. Annealing}
While quantum annealing (e.g., D-Wave systems) is often used for QUBO problems, it is restricted to specific physical architectures and cannot execute general-purpose quantum algorithms. In this research, we exclusively employ gate-based universal quantum computers. Gate-based models, such as those from IBM and Google, offer greater algorithmic flexibility, allowing for the precise implementation of parameterized quantum circuits, entanglement layers, and the seamless integration of CVaR-based sampling. This universality is a prerequisite for exploring adaptive, multi-layered variational architectures and comparing various classical optimizers within a unified framework.

\subsection{Proposed Adaptive Layer-CVaR-VQE}
Building upon the fundamental concepts of the VQE and the RNA Hamiltonian, this section introduces the proposed Layer-CVaR-VQE architecture. We first detail how the Layer-VQE framework dynamically increases the circuit depth to capture complex structural correlations, followed by an explanation of the Conditional Value-at-Risk (CVaR) objective function, which refines the sampling efficiency. Finally, we discuss the critical role of selecting an appropriate classical optimizer to navigate the resulting non-convex energy landscape and formally establish the first adaptive quantum approach for RNA folding.

\subsubsection{Layer VQE Framework}
Standard VQE architectures utilize a fixed-depth ansatz, which may fail to capture the complex correlations in RNA structures or might introduce unnecessary noise if the depth exceeds the problem's requirement. To address this, we adopt the Layer-VQE concept of \citet{liu2022layer}, which we apply here to RNA folding for the first time. The Layer-VQE framework progressively increases the circuit depth by adding parameterized rotation blocks and entangling layers. 

We initialize the circuit with single-qubit \(R_y\) gates to create superposition, followed by entangling layers (e.g., CNOT gates) to generate correlations. The number of layers \(L\) determines the expressivity of the ansatz. The total number of trainable parameters scales linearly as \(L \times Q\). As \(L\) increases, the circuit explores a larger Hilbert space. However, excessive depth in the NISQ era leads to barren plateaus and noise accumulation. Therefore, our approach evaluates the energy at each layer independently to find the optimal structural correlation without over-parameterization. By extracting the best-performing layer, the algorithm adapts its depth to the specific structural complexity of the given RNA sequence.

\begin{figure}[!htbp]
    \centering
    \includegraphics[width=0.75\columnwidth]{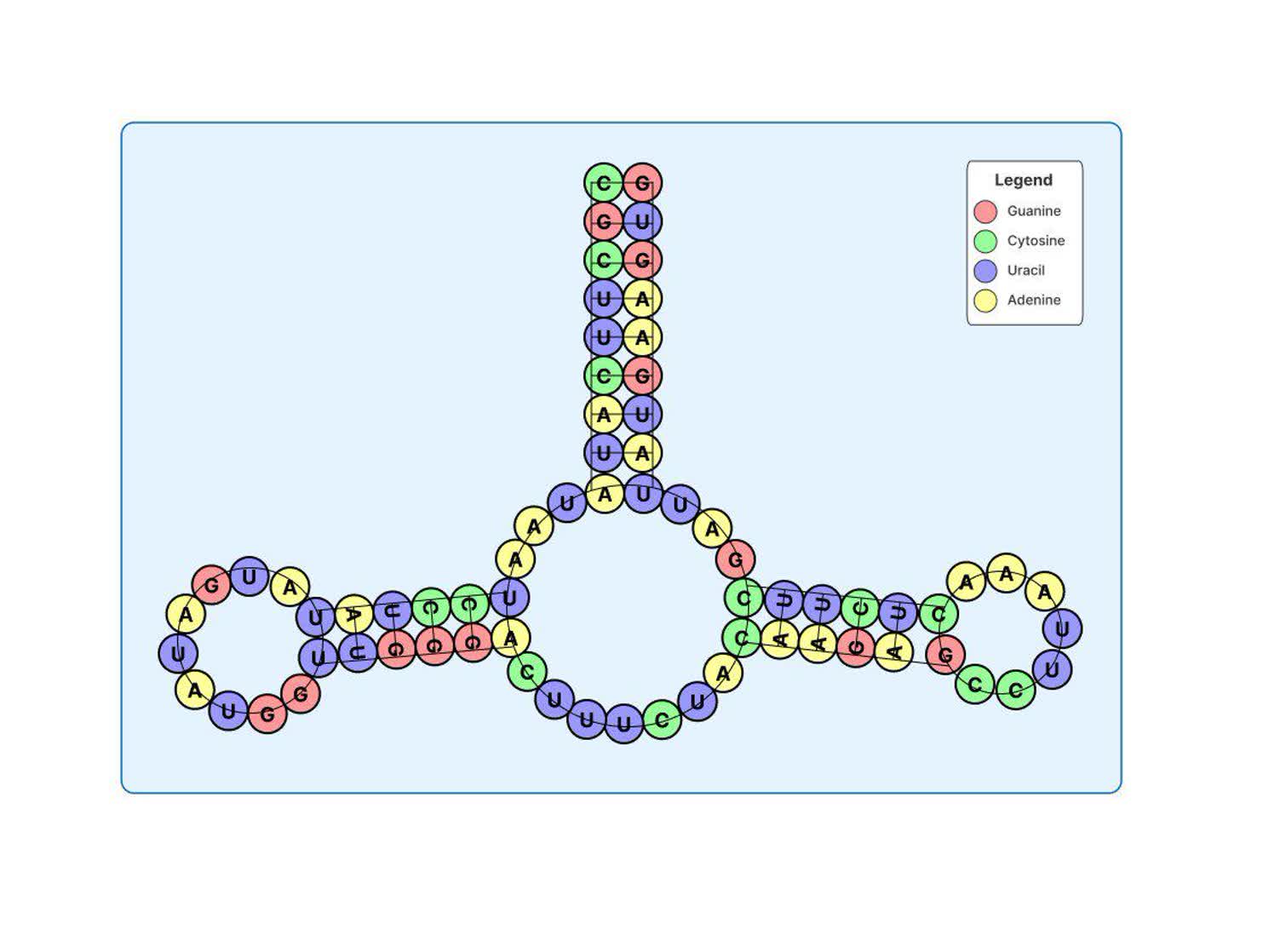}\vspace{-8pt}
    \caption{Visualization of a folded RNA secondary structure, highlighting the canonical base pairings within a stem and a hairpin loop. The structure was drawn with an online visualization tool.}
    \label{fig:rna_example}
\end{figure}

To better understand the biological target of this optimization, Figure~\ref{fig:rna_example} presents a conceptual example of a folded RNA molecule. The structure features a canonical stem formed by Watson-Crick base pairings and a hairpin loop, which are the primary motifs the Hamiltonian is designed to identify and reward.
\subsubsection{Layer VQE Procedure}
Algorithm~\ref{alg:lvqe} summarizes the Layer VQE procedure of \citet{liu2022layer}, which we adapt to the RNA folding Hamiltonian. The algorithm starts from a shallow ansatz with one \(R_y\) rotation on each qubit and optimizes it for a fixed budget of \(k_0\) iterations. It then adds one layer at a time. Each new layer is initialized to the identity, so the energy does not jump when the layer is inserted, and the parameters of all layers are re-optimized for \(k_\ell\) iterations. After the last layer is added, a final optimization run over all parameters is carried out until convergence.

Two points matter for our use of the procedure. First, the early stop at \(k_\ell\) keeps the cost of each depth low, so the search never spends the full budget on a circuit that is too deep for the sequence. Second, starting each new layer at the identity means that the deeper circuit inherits the solution already found at the previous depth. In our experiments, the energy reached at layer \(\ell\) is therefore used to choose the depth, as discussed in the Results section.
\begin{algorithm}[!htbp]
\caption{L-VQE with $\ell$ layers, adapted from \citet{liu2022layer}.}
\label{alg:lvqe}
\begin{algorithmic}[1]
\STATE Initialize the ansatz with one $R_y$ acting on each qubit.
\STATE Update the parameters to minimize $\langle\psi_0(\boldsymbol{\theta})|\mathcal{H}|\psi_0(\boldsymbol{\theta})\rangle$; stop after $k_0$ iterations (before reaching convergence).
\FOR{$l=1,\dots,\ell$}
\STATE Add a new layer to the ansatz and initialize it such that it evaluates to identity.
\STATE Update all parameters to minimize $\langle\psi_l(\boldsymbol{\theta})|\mathcal{H}|\psi_l(\boldsymbol{\theta})\rangle$; stop after $k_\ell$ iterations (before reaching convergence).
\ENDFOR
\STATE Update all parameters to minimize $\langle\psi(\boldsymbol{\theta})|\mathcal{H}|\psi(\boldsymbol{\theta})\rangle$ until convergence.
\end{algorithmic}
\end{algorithm}
\subsubsection{Conditional Value-at-Risk (CVaR) Objective}
To improve the sampling efficiency, we employ the CVaR-VQE variant \citep{barkoutsos2019improving}. Instead of minimizing the expected value of the energy over all measured bitstrings, CVaR-VQE focuses only on the lowest-energy tail of the distribution. If we measure the system \(N\) times, we sort the resulting energies \(E_1 \le E_2 \le \dots \le E_N\). The CVaR objective with a confidence level \(\alpha \in (0,1]\) is defined as the average of the lowest \(\alpha\) fraction of the samples:
\begin{equation}
    \text{CVaR}_\alpha = \frac{1}{\lceil \alpha N \rceil} \sum_{k=1}^{\lceil \alpha N \rceil} E_k
\end{equation}
By setting \(\alpha = 0.2\), the optimizer focuses on the top 20\% most stable conformations, accelerating convergence toward the global minimum energy. 

\subsubsection{The Role of Optimizer Selection}
The landscape of the CVaR objective function in the NISQ era is highly non-convex, discrete, and noisy (due to finite shots). Classical optimizers interact differently with this landscape. Gradient-based methods often fail due to the non-differentiability of the bitstring sampling. Population-based and derivative-free algorithms can explore the space more robustly but may suffer from slow convergence. Because previous studies typically relied on a single optimizer (e.g., NFT or SPSA), the true impact of optimizer selection on RNA folding remained unexplored. To bridge this gap, we benchmark seven algorithms—GA, PSO, PS, SA, SUR, SPSA, and COBYLA—to identify which class of optimizers is best suited for navigating the quantum RNA folding landscape.

\subsubsection{The First Adaptive Approach for RNA}
This work marks the first application of an adaptive layered quantum approach to the RNA secondary structure prediction problem. We adopt the Layer VQE procedure of \citet{liu2022layer}, who showed that incrementally deepening the circuit helps on noisy hardware and reduces the risk of poor local minima. Instead of relying on a fixed circuit depth, the Layer-CVaR-VQE method allows the quantum circuit to grow layer by layer, evaluating the energy at each step. This adaptive behavior is crucial because different RNA sequences exhibit varying degrees of structural complexity (quantified by \(Q\)). For sequences with a small number of candidate pairs, a shallow circuit (Layer 1 or 2) is sufficient. For complex sequences with large \(Q\), deeper layers are necessary to capture long-range interactions. By selecting the best-performing layer dynamically, the algorithm balances expressivity and circuit noise, significantly outperforming the standard single-layer CVaR-VQE.

\section{Results and Discussion}
\label{sec:results}
This section first describes the ansatz and how the sequence is encoded into qubits. It then reports the performance of the seven optimizers against the three baselines, followed by the statistical tests and the layer-wise analysis. A single-sequence case study closes the section.
\begin{figure}[!htbp]
    \centering
    \includegraphics[width=0.6\columnwidth]{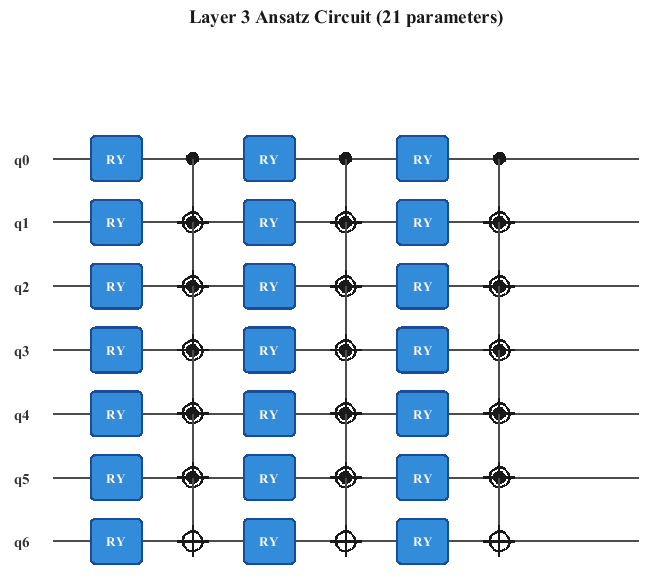}
    \caption{Three-layer hardware-efficient ansatz (21 parameters) used in the main benchmark experiments, shown for $Q=7$ qubits. Each layer consists of $R_y$ rotations on all qubits followed by CNOTs from $q_0$ to $q_1$--$q_6$.}
    \label{fig:ansatz}
\end{figure}
\subsection{Example Ansatz Circuit}
The ansatz used in all benchmark experiments reported below (Figures~\ref{fig:overall_sens}--\ref{fig:best_layer}) is the three-layer circuit shown in Figure~\ref{fig:ansatz}, drawn here for $Q=7$ qubits. Each layer applies a parameterized $R_y(\theta)$ rotation to every qubit $q_0,\dots,q_6$ (7 parameters per layer, 21 in total), followed by an entangling block in which $q_0$ is the control of CNOT gates targeting $q_1,\dots,q_6$. Since these CNOTs share the same control they commute, so the fan-out in the figure is equivalent to a sequence of individual CNOT gates.

To make the encoding concrete, Table~\ref{tab:qubit_map} lists the qubit-to-base-pair mapping for the sequence CGGAAUCCGG, which is also used in the case study of the next subsection.

\begin{table}[!htbp]
\centering
\caption{Qubit encoding of the candidate base pairs for CGGAAUCCGG ($N=10$, $Q=7$). A bitstring bit equal to 1 means the pair is formed.}
\label{tab:qubit_map}
\begin{tabular}{|c|c|c|}
\hline
\textbf{Qubit (bit)} & \textbf{Base pair $(i,j)$} & \textbf{Nucleotides} \\
\hline
1 & (1,9) & C--G \\
2 & (1,10) & C--G \\
3 & (2,7) & G--C \\
4 & (2,8) & G--C \\
5 & (3,7) & G--C \\
6 & (3,8) & G--C \\
7 & (7,10) & C--G \\
\hline
\end{tabular}
\end{table}

\subsection{Case Study: Choosing the Circuit Depth for CGGAAUCCGG with PSO}
This subsection is a separate single-sequence illustration and is not part of the benchmark statistics reported elsewhere; here the circuit was run with five layers (35 parameters) to show how the choice of depth affects the decoded structure of one sequence.

\begin{figure*}[!t]
    \centering
    \includegraphics[width=0.95\textwidth]{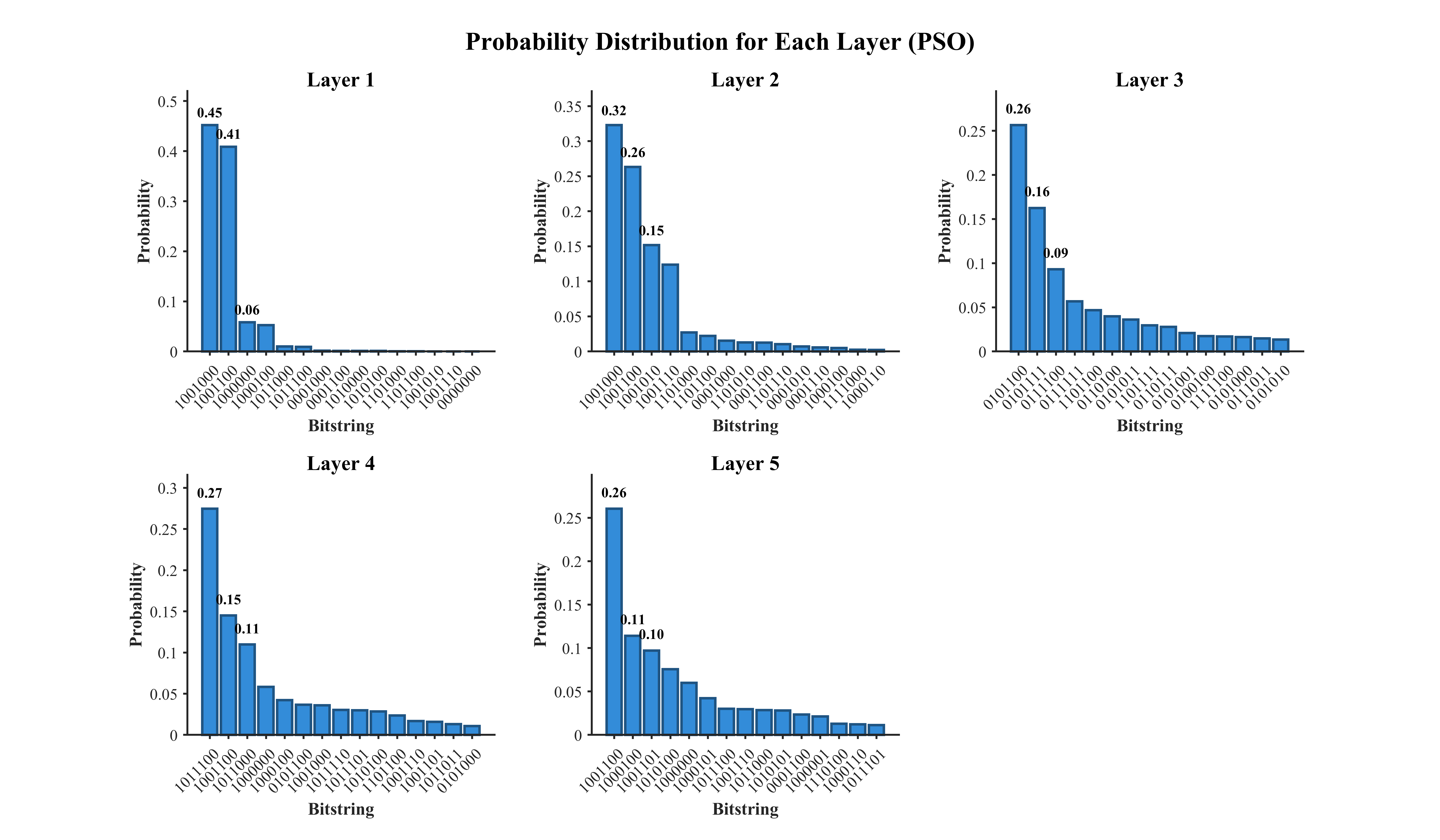}
    \caption{Probability distribution of the bitstrings for each layer (Layers 1--5) of the PSO-optimized Layer-CVaR-VQE for CGGAAUCCGG. The target bitstring 1001100 is the most probable state only in Layer 5.}
    \label{fig:pso_layers}
\end{figure*}

Figure~\ref{fig:pso_layers} shows the probability distribution of each of the five layers for CGGAAUCCGG optimized by PSO, using the seven-qubit encoding of Table~\ref{tab:qubit_map}. The target bitstring 1001100 corresponds to the pairs (1,9), (2,8), and (3,7), that is, the dot-bracket structure \texttt{(((...))).}, which agrees with ViennaRNA and ProbKnot. Layer selection is based on two criteria: whether the target bitstring is decoded correctly, and the margin between its probability and that of the next most probable bitstring.

In Layer 1, the correct bitstring 1001100 is second with 41\%, behind the folded bitstring 1001000 (45\%), which encodes only the pairs (1,9) and (2,8). In Layer 2, 1001100 is second with 26\%, behind 1001000 (32\%). In Layer 3, the target bitstring does not appear among the highest-probability states. In Layer 4, 1001100 is second with 15\%, behind 1011100 (27\%). Layer 5 is the only layer in which the target bitstring is the most probable state, with 26\%. The next most probable state, 1000100, has 11\%, a margin of 15 percentage points. Layer 5 is therefore the most suitable depth for this sequence under PSO.

\subsection{Simulation Environment and Hyperparameter Configuration}
The proposed Layer-CVaR-VQE framework was implemented using the MATLAB Quantum Computing Toolbox (R2024a). The simulations were conducted on a statevector-based quantum simulator, allowing for noiseless execution of the parameterized quantum circuits. To ensure a rigorous evaluation, a benchmark dataset comprising 23 RNA sequences was compiled. The lengths of these sequences range from 9 to 25 nucleotides, resulting in a search space of \(Q\) valid candidate base pairs ranging from 6 to 18 qubits. 

The performance of seven classical optimizers was benchmarked within the variational loop. The hyperparameters for each optimizer were carefully selected to balance exploration and computational budget, as detailed in Table~\ref{tab:optimizers}. For all experiments, the CVaR confidence level was fixed at \(\alpha = 0.2\), and the quantum state was sampled using 1000 shots per evaluation. 

As outlined in Table~\ref{tab:optimizers}, the hyperparameters for each classical optimizer were carefully configured to balance exploration and the computational budget. For the Genetic Algorithm (GA), a population size of 20 was used with a crossover fraction of 0.8 and an elite count of 2, allowing it to retain the best solutions while maintaining diversity. Particle Swarm Optimization (PSO) was configured with a swarm size of 4, which is relatively small but sufficient for the limited parameter space of the layered ansatz. The SPSA algorithm employed standard decay parameters (a=0.2, c=0.1, \(\alpha\)=0.602, \(\gamma\)=0.101) to balance gradient approximation and convergence speed. For the remaining optimizers, including Pattern Search, Simulated Annealing, Surrogate, and COBYLA, the tolerances and evaluation limits were set to ensure a fair comparison under the same computational budget of 20 to 200 function evaluations.

\begin{table*}[!htbp]
\centering
\caption{Hyperparameter Configuration and Parameter Descriptions for the Seven Benchmark Optimizers}
\label{tab:optimizers}
\vspace{8pt}
\renewcommand{\arraystretch}{1.3}
\setlength{\tabcolsep}{6pt}
\begin{tabular}{|l|p{7.5cm}|c|}
\hline
\textbf{Algorithm} & \textbf{Key Hyperparameters (Description)} & \textbf{Max Evaluations} \\ \hline
\textbf{GA} & Population Size = 20, Crossover Fraction = 0.8, Elite Count = 2 & 20 Generations \\ \hline
\textbf{PSO} & Swarm Size = 4, Inertia Weight (default) & 20 Iterations \\ \hline
\textbf{PS} & Mesh Tolerance = \(10^{-4}\) (Mesh resolution limit), Step Tolerance = \(10^{-4}\) (Minimum step size) & 20 Iterations \\ \hline
\textbf{SPSA} & \(a = 0.2\) (Step size), \(c = 0.1\) (Perturbation size), \(\alpha = 0.602\) (Step decay), \(\gamma = 0.101\) (Perturbation decay) & 20 Iterations \\ \hline
\textbf{SA} & Initial Temperature = 100, Reanneal Interval = 50 (Cooling steps) & 20 Iterations \\ \hline
\textbf{SUR} & Initial Points = Random (Sampling strategy) & 30 Evaluations \\ \hline
\textbf{COBYLA} & Interior-point, Tolerance = \(10^{-6}\) (Convergence threshold) & 200 Evaluations \\ \hline
\end{tabular}
\vspace{0.3cm} 
\end{table*}
\begin{figure*}[!htbp]
    \centering
    \includegraphics[width=\textwidth]{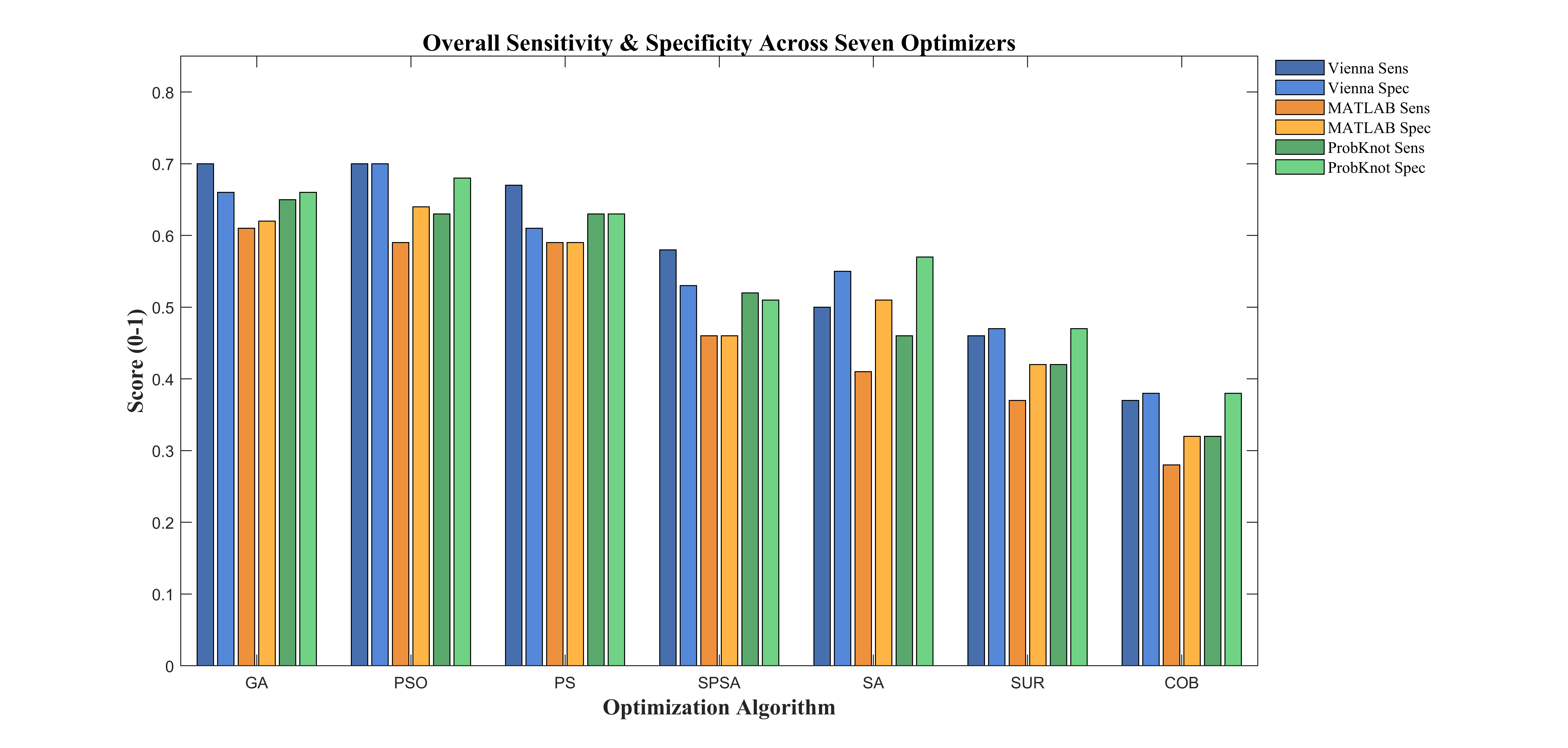}
    \caption{Overall sensitivity and specificity comparison of seven optimization algorithms within the Layer-CVaR-VQE framework against three classical baselines.}
    \label{fig:overall_sens}
\end{figure*}
\subsection{Overall Performance of Optimization Algorithms}
Figure~\ref{fig:overall_sens} presents the overall sensitivity and specificity of the seven optimizers evaluated against three widely used classical baseline methods: ViennaRNA, MATLAB MFE, and ProbKnot. It is immediately evident that the nature-inspired and population-based algorithms significantly outperform the gradient-dependent and surrogate-based methods. 

The Genetic Algorithm (GA) achieves the most balanced performance, yielding a mean sensitivity of 70.0\% and a specificity of 66.0\% against ViennaRNA. Particle Swarm Optimization (PSO) follows closely with a mean sensitivity of 70.0\% and a notably high specificity of 70.0\%. Pattern Search (PS) ranks third, maintaining consistent scores of 67.0\% sensitivity and 61.0\% specificity across all three reference baselines. Conversely, COBYLA exhibits the weakest performance, with its sensitivity dropping below 37.0\%. This substantial performance gap (over 30\% difference between GA and COBYLA) highlights the inadequacy of gradient-based methods for the highly non-convex, discrete, and shot-noisy cost landscape of the RNA folding Hamiltonian.

\subsection{Distribution Analysis and Statistical Significance}
\begin{figure*}[!htbp]
    \centering
    \includegraphics[width=0.9\textwidth]{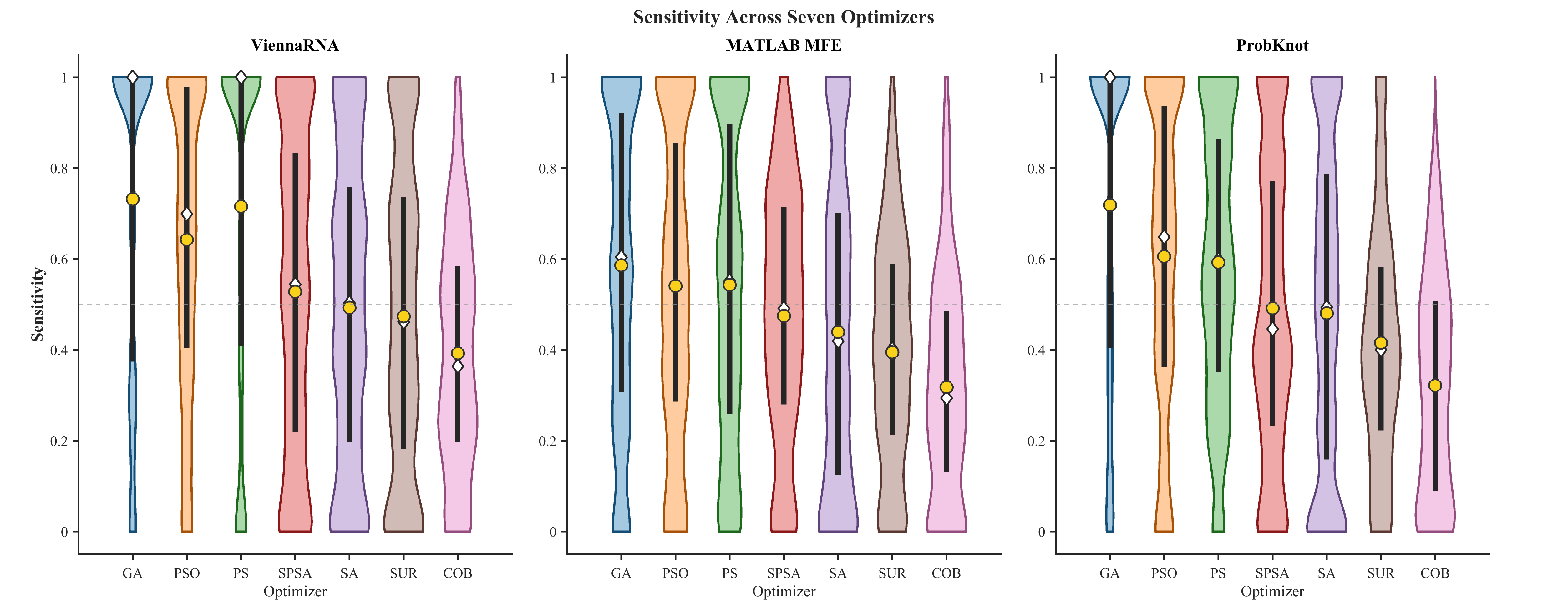}
    \caption{Combined violin distributions of sensitivity across seven optimizers evaluated against ViennaRNA, MATLAB MFE, and ProbKnot.}
    \label{fig:violin_sens}
\end{figure*}

\begin{figure*}[!htbp]
    \centering
    \includegraphics[width=0.9\textwidth]{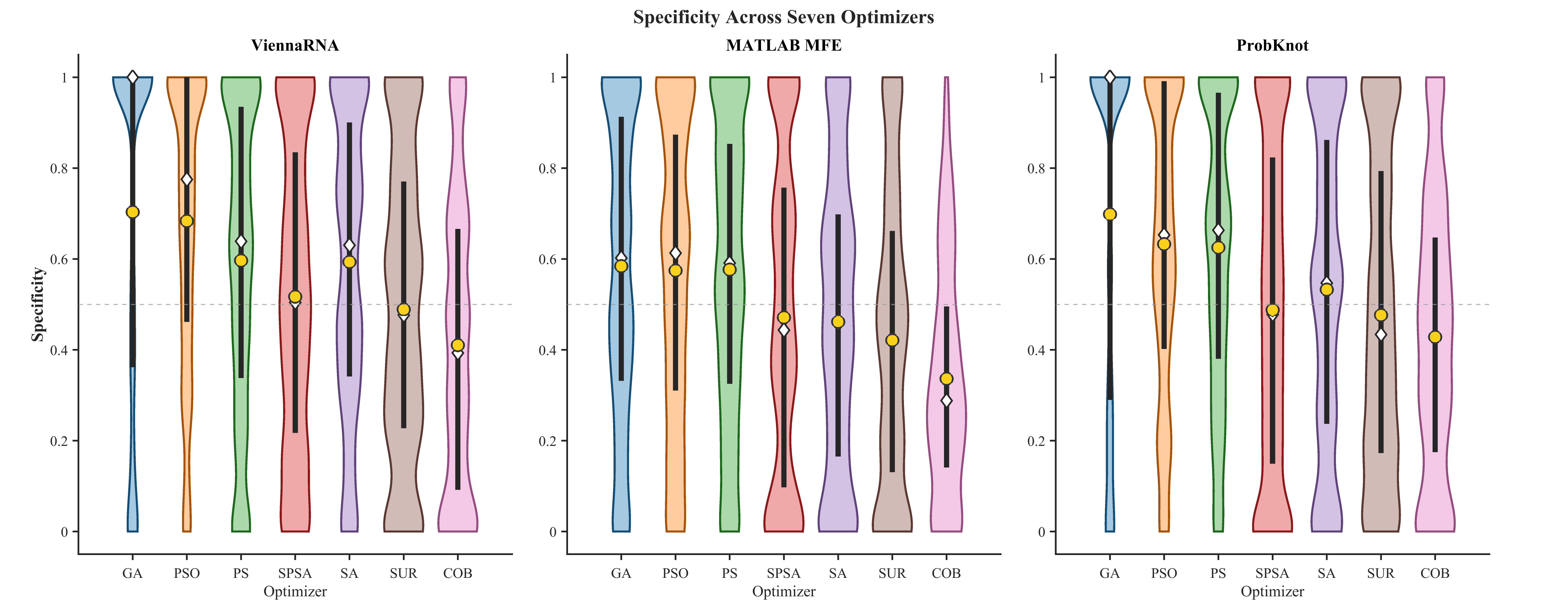}
    \caption{Combined violin distributions of specificity across seven optimizers evaluated against ViennaRNA, MATLAB MFE, and ProbKnot.}
    \label{fig:violin_spec}
\end{figure*}

While the mean scores provide a baseline comparison, the distribution of results across the 23 sequences reveals critical insights into algorithmic stability. Figures~\ref{fig:violin_sens} and \ref{fig:violin_spec} display the combined violin plots for sensitivity and specificity. 

The violin plots demonstrate that GA and PSO possess the narrowest interquartile ranges and the highest medians, indicating robust and consistent convergence. Notably, GA exhibits a bimodal distribution for sensitivity against ViennaRNA, with a median reaching 1.00. This indicates that for more than half of the test sequences, GA recovers the exact reference base-pairing pattern. In contrast, the lower-performing algorithms (SPSA, SA, Surrogate, and COBYLA) display broad, low-centered distributions, reflecting their tendency to become trapped in local minima. The MATLAB MFE baseline consistently yielded the lowest scores across all optimizers, primarily because its minimum free-energy predictions tend to under-predict base pairs for short RNA fragments, thereby inflating the false-negative rate.

The superior stability of GA can be attributed to its inherent evolutionary mechanisms. By maintaining a diverse population and employing crossover and mutation, GA avoids the premature convergence that plagues single-solution optimizers. Similar observations have been reported in recent hybrid quantum-classical literature. For instance, Zaborniak et al. \citep{zaborniak2022qubo} noted the challenges of local minima when using SPSA for QUBO formulations of RNA. While their approach relied on a single trajectory, our population-based approach effectively samples multiple regions of the Hilbert space simultaneously, resulting in a more reliable exploration of the energy landscape.

Furthermore, PSO demonstrates a similarly robust profile. PSO's swarm intelligence allows particles to share information about the best-found solutions, dynamically updating their velocities to converge toward optimal bitstrings. This mechanism is particularly well-suited for the shot-noisy objective function of CVaR-VQE, where the energy evaluation fluctuates between iterations. Alevras et al. \citep{alevras2024mrna} highlighted the sensitivity of NFT to noise during their utility-scale demonstrations. Our results suggest that gradient-free metaheuristics like PSO can offer a more resilient alternative in the presence of finite sampling noise, albeit sometimes at the cost of more function evaluations.

The intermediate algorithms, such as Pattern Search and Simulated Annealing, exhibit a trade-off. Pattern Search provides a deterministic local refinement that is highly effective when the initial point is close to the global minimum. This is consistent with its performance against MATLAB MFE, where it performs competitively. However, its inability to accept worse solutions makes it prone to getting stuck in the rugged, non-convex barriers of the RNA landscape. Simulated Annealing, on the other hand, can probabilistically escape local minima. Yet, with only 20 iterations, the cooling schedule is too rapid to fully exploit this advantage, leading to the wide, low-centered distributions observed in Figures~\ref{fig:violin_sens} and \ref{fig:violin_spec}.

Finally, the Surrogate and COBYLA methods exhibit the highest variance. COBYLA is fundamentally mismatched to this problem because it approximates gradients based on finite differences. In a landscape where the objective function changes discretely with bitstring evaluations and is subject to shot noise, numerical gradient estimation is highly unreliable. This leads to erratic trajectories and convergence failures \citep{shaffer2023surrogate}, as evidenced by the lowest medians and broadest spans. This aligns with the theoretical limitations of derivative-based methods for combinatorial optimization on NISQ devices, as discussed by Oliv et al. \citep{oliv2022evaluating}.

\begin{figure}[!htbp]
    \centering
    \includegraphics[width=\columnwidth]{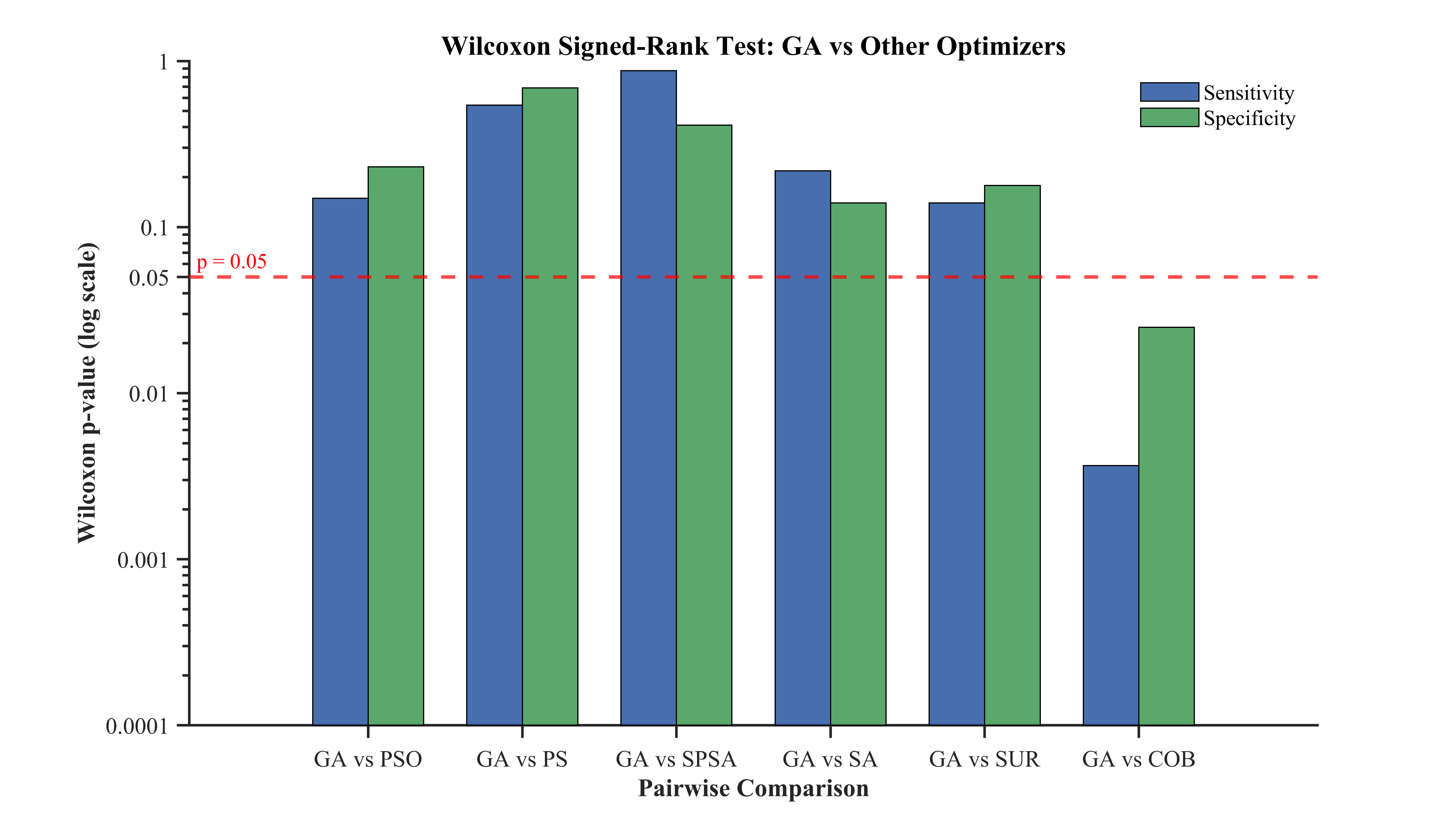}
    \caption{Wilcoxon signed-rank test p-values for sensitivity and specificity, comparing GA against other optimizers.}
    \label{fig:wilcoxon}
\end{figure}

To statistically validate the superiority of the top-performing algorithm, a Wilcoxon signed-rank test was conducted, comparing GA against the other six optimizers. As shown in Figure~\ref{fig:wilcoxon}, the p-values for the pairwise comparisons (GA vs. PSO, PS, SPSA, SA, SUR, COBYLA) are reported per metric. Pairs with $p<0.05$ indicate a statistically significant GA advantage; pairs above the threshold indicate no significant difference from GA. The test confirms that the performance gap observed in the visual distributions is rooted in statistically significant differences across the benchmark dataset.

\begin{figure*}[!htbp]
    \centering
    \includegraphics[width=0.8\textwidth]{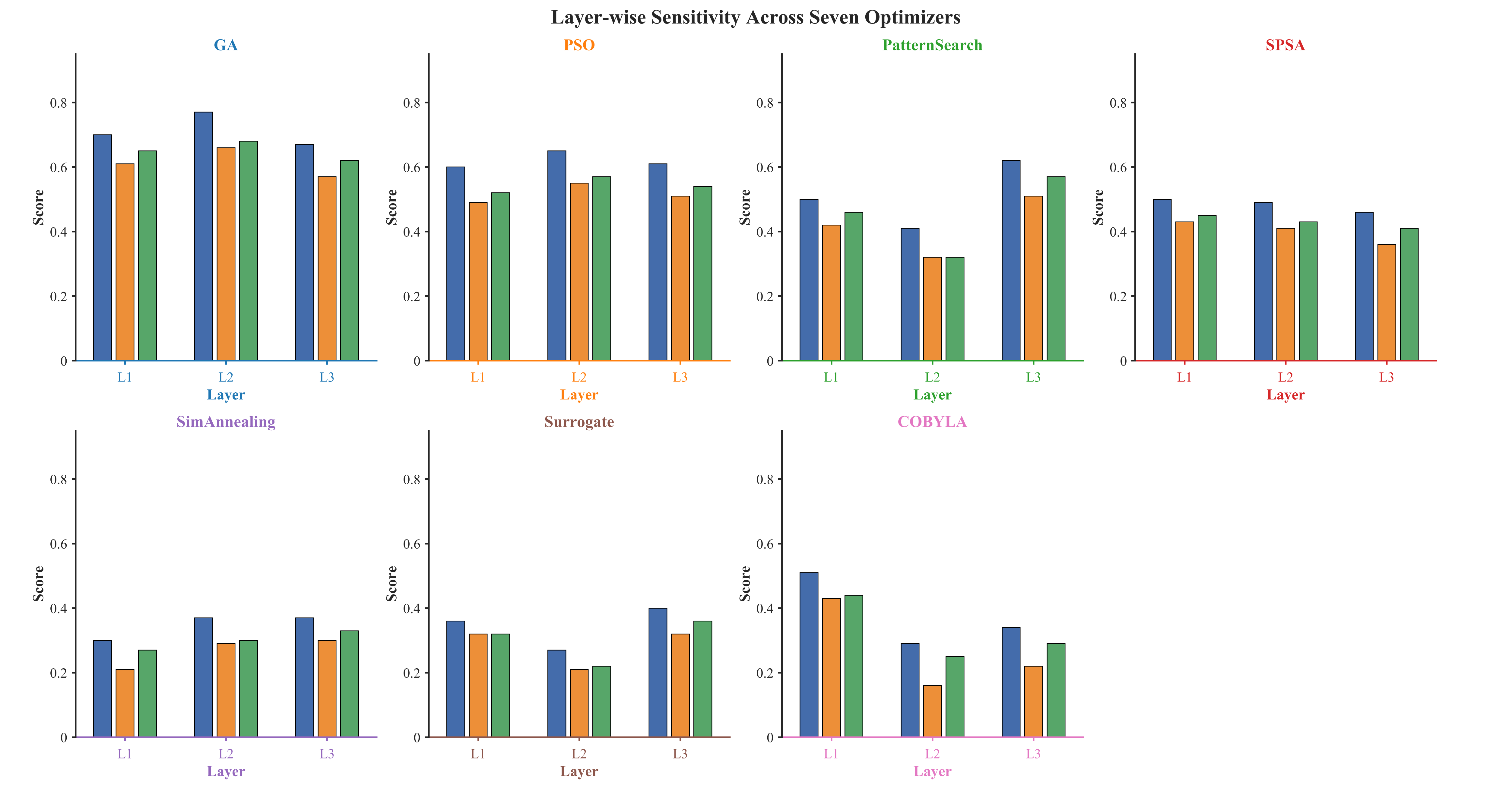}
    \caption{Layer-wise sensitivity comparison across seven optimizers. Intermediate depths (L2 and L3) consistently outperform the standard single-layer (L1) approach.}
    \label{fig:layer_sens}
\end{figure*}

\begin{figure*}[!htbp]
    \centering
    \includegraphics[width=0.8\textwidth]{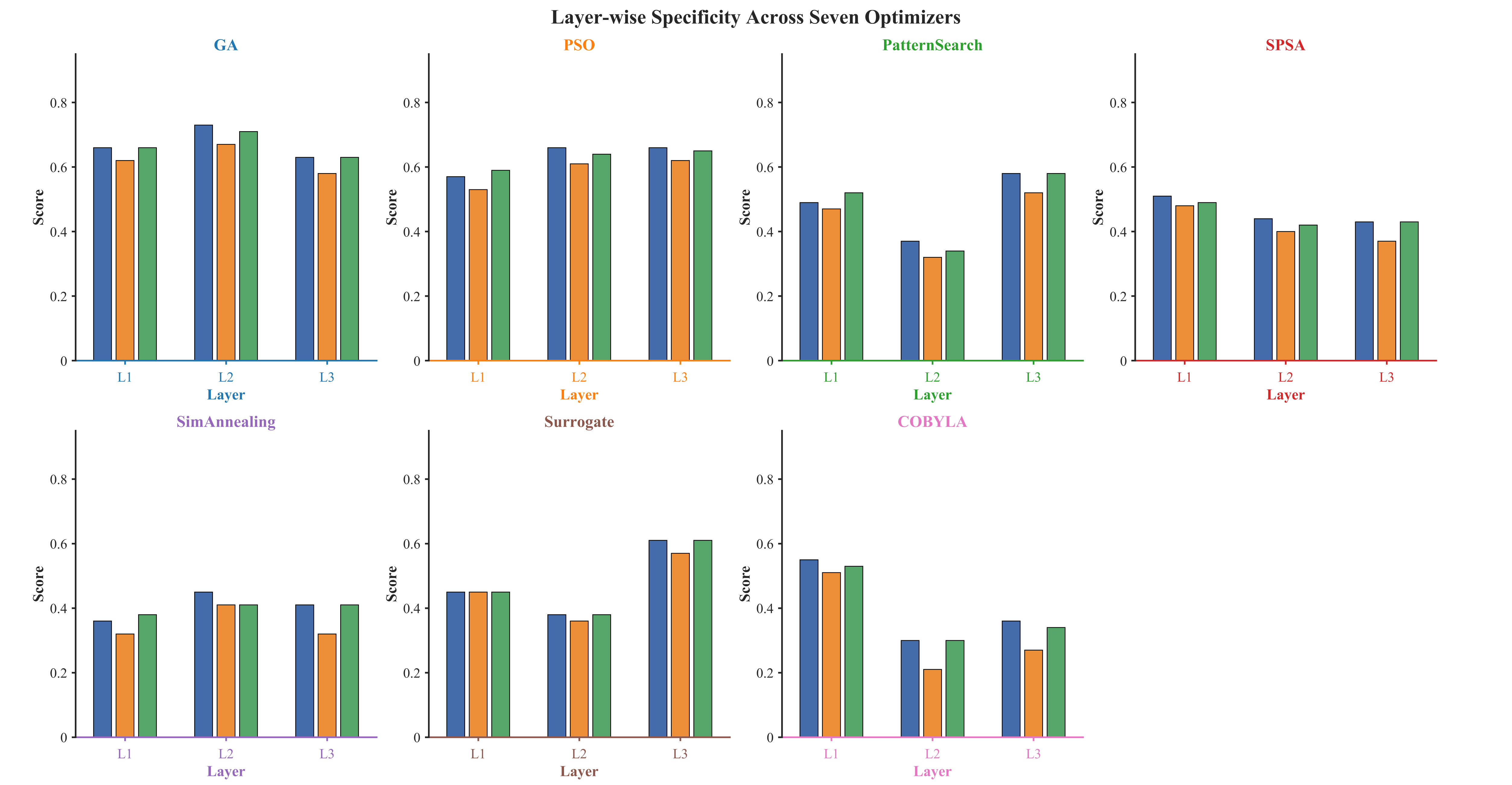}
    \caption{Layer-wise specificity comparison across seven optimizers, demonstrating the effectiveness of the adaptive layered ansatz.}
    \label{fig:layer_spec}
\end{figure*}

\subsection{Layer-wise Analysis: Validating the Adaptive Approach}
A central hypothesis of this research is that the standard CVaR-VQE (which corresponds to a single-layer ansatz) is insufficient for complex RNA structures. By expanding the circuit depth, the Layer-CVaR-VQE architecture allows the quantum state to explore a larger Hilbert space. Figures~\ref{fig:layer_sens} and \ref{fig:layer_spec} illustrate the layer-wise sensitivity and specificity for all seven optimizers across Layer 1 (L1), Layer 2 (L2), and Layer 3 (L3). The results strongly validate the adaptive layering strategy. For the vast majority of optimizers, the intermediate layers (L2 or L3) significantly outperform the first layer. For instance, in the GA model, the mean sensitivity against ViennaRNA increases from 0.70 at L1 to 0.77 at L2, while specificity rises from 0.66 to 0.73. This represents a substantial accuracy improvement (approximately 7\%) solely by increasing the circuit depth to the second layer.

The underlying reason for this improvement is deeply rooted in the entanglement capabilities of the quantum circuit. In L1, the circuit only applies a single layer of \(R_y\) rotations and linear CNOTs. While this is sufficient to create basic superpositions, it struggles to encode the long-range correlations necessary for identifying complex RNA motifs, such as nested stems separated by large loops. As the circuit depth reaches L2 and L3, the additional entangling layers allow the qubits to interact more thoroughly, effectively correlating distant base pairs within the Hamiltonian. This enhanced expressivity aligns with the theoretical framework of the Layer-VQE concept proposed by Liu et al. \citep{liu2022layer}, which states that intermediate circuit depths are crucial for navigating complex combinatorial landscapes.

\begin{figure*}[!htbp]
    \centering
    \includegraphics[width=0.8\textwidth]{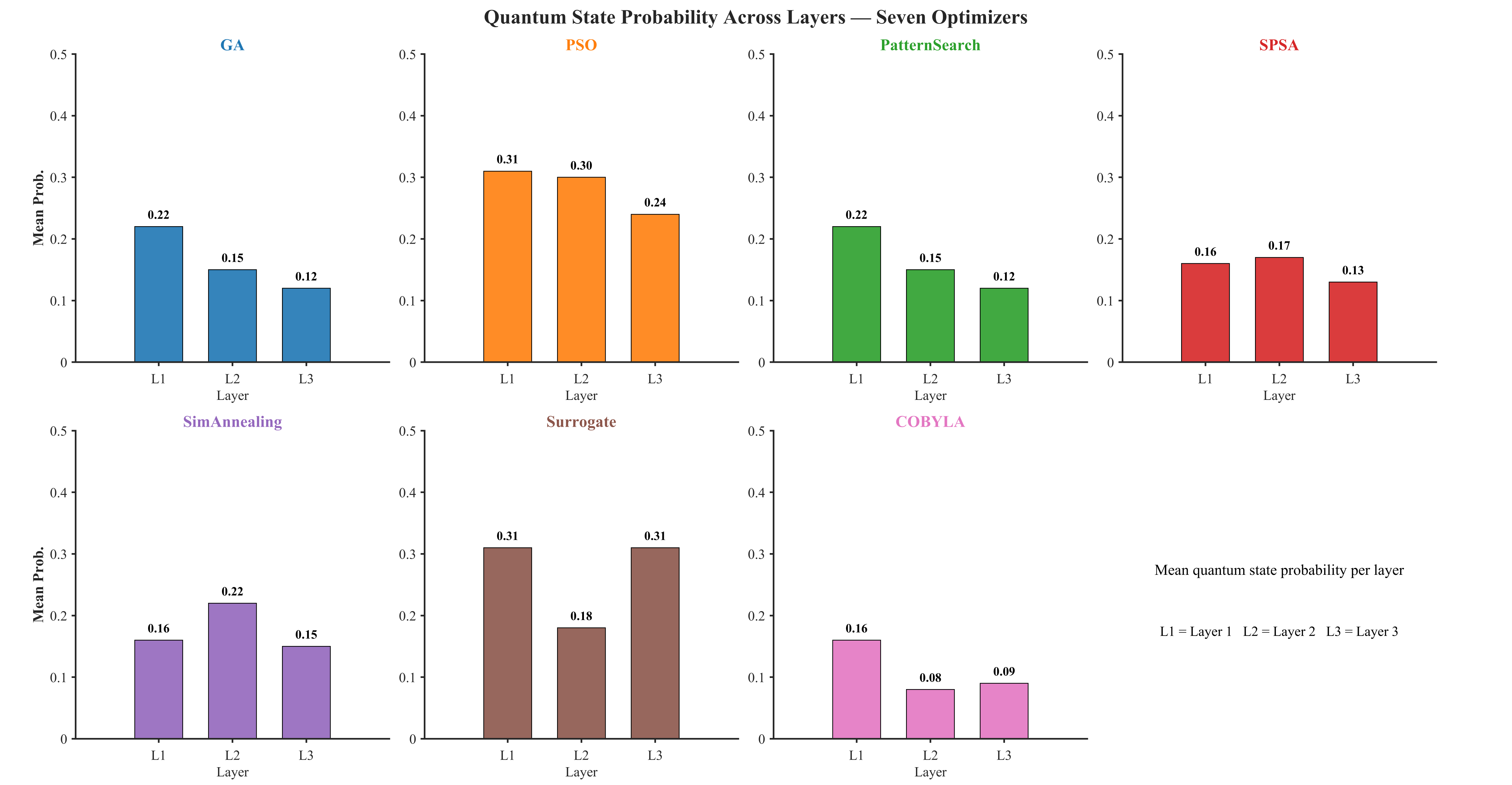}
    \caption{Mean quantum state probability per layer for all seven optimizers, illustrating amplitude concentration across circuit depths.}
    \label{fig:prob_layer}
\end{figure*}

However, the benefits of deeper circuits are not infinite. Our observations indicate that performance frequently degrades beyond L3. This phenomenon is consistent with the onset of "barren plateaus" and the accumulation of gate errors in the NISQ era. When the number of parameters \(L \times Q\) grows too large, the optimization landscape flattens exponentially, making it difficult for any classical optimizer to find the global minimum. Furthermore, the limited number of shots (1000 per evaluation) introduces statistical noise that compounds with circuit depth. Therefore, the optimal depth represents a delicate balance between the required expressivity to capture RNA topology and the circuit's susceptibility to noise and optimization stagnation.

The layer-wise results also highlight an interesting interaction between the optimizer's characteristics and the circuit depth. GA, with its robust global search, successfully extracts value from the deeper layers. In contrast, Pattern Search shows its most dramatic improvement at L3, likely because its deterministic mesh-size refinement benefits from the larger parameter space to pinpoint the exact ground state after the global structure is roughly located. On the other hand, COBYLA's performance consistently deteriorates as depth increases, confirming its inadequacy in handling the high-dimensional, noisy parameter spaces introduced by deeper ansatzes. These findings underscore that algorithm-circuit co-design is essential for RNA folding, as a fixed-depth architecture cannot universally accommodate the diverse topological needs of different optimizers.

\subsection{Quantum State Probability and Layer Selection Dynamics}
The concentration of probability on the optimal basis state serves as a direct indicator of the algorithm's confidence in its structural prediction. Figure~\ref{fig:prob_layer} displays the mean quantum state probability for each layer across all optimizers. A higher probability mass on the target bitstring implies that the variational state has successfully localized its amplitude, increasing the likelihood of sampling the correct secondary structure. The probability mass is generally highest at L1 and L2, which corroborates the accuracy trends observed in the previous section.

PSO achieves the highest mean probability at L1 (31.0\%), while Surrogate and GA maintain strong probabilities at L2 (31.0\% and 15.0\%, respectively). Interestingly, while Surrogate occasionally reaches 100\% probability, it converges entirely on the all-zero bitstring, representing a completely unfolded RNA. This serves as a critical warning: high probability does not guarantee structural accuracy if the optimizer lacks global search capabilities. The objective landscape may have a local minimum at the all-zero state if the penalty for forming pairs initially outweighs the rewards, trapping gradient-descent-like algorithms in a trivial solution.

\begin{figure*}[!htbp]
    \centering
    \includegraphics[width=0.9\textwidth]{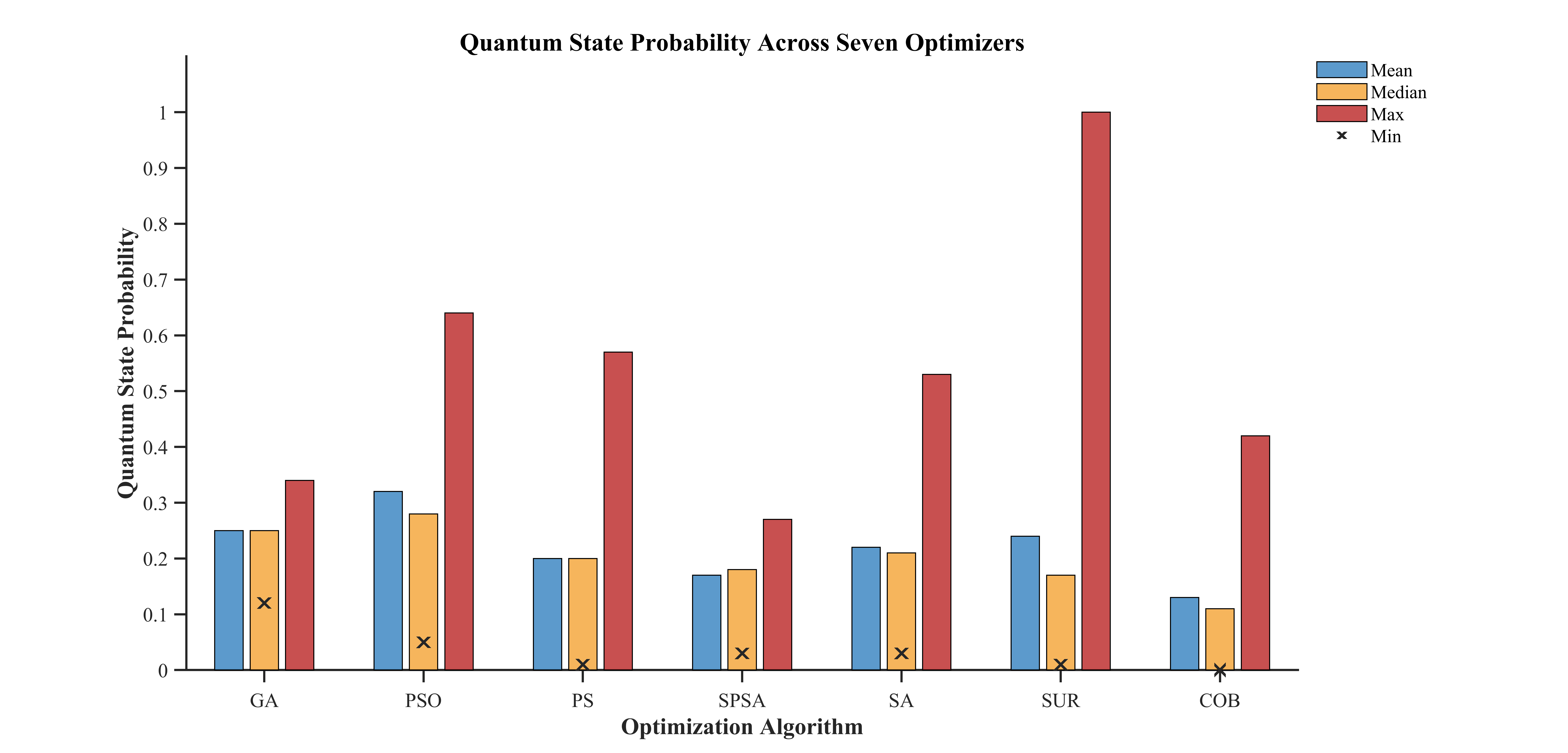}
    \caption{Overall quantum state probability statistics (mean, median, max, and min) across the seven optimizers.}
    \label{fig:prob_algs}
\end{figure*}

From a quantum information perspective, the CVaR objective naturally drives the amplitude concentration toward the lowest-energy states. However, the ability to localize this amplitude is heavily dependent on the optimizer. Gradient-based optimizers, such as SPSA, often experience oscillatory behavior in the probability mass. Because they rely on finite difference approximations of the gradient, the inherent shot noise of the quantum measurement process can misdirect the parameter updates. As a result, SPSA fails to consistently concentrate the amplitude, leading to a broader distribution of probabilities across suboptimal states.

The overall probability statistics (Figure~\ref{fig:prob_algs}) reveal that GA and PSO achieve the most stable and highest mean probabilities (24.5\% and 32.5\%), corresponding to their superior convergence profiles. GA's evolutionary approach ensures that even if some members of the population are trapped in a local minimum, the global population retains diversity, pushing the amplitude toward the true ground state. In contrast, COBYLA maintains the lowest mean probability (13.0\%), reflecting its inability to drive the quantum amplitude toward the optimal states. This confirms that population-based methods are superior at navigating the noisy Hilbert space and concentrating the probability amplitude on the biologically meaningful conformations.

\begin{figure}[!htbp]
    \centering
    \includegraphics[width=\columnwidth]{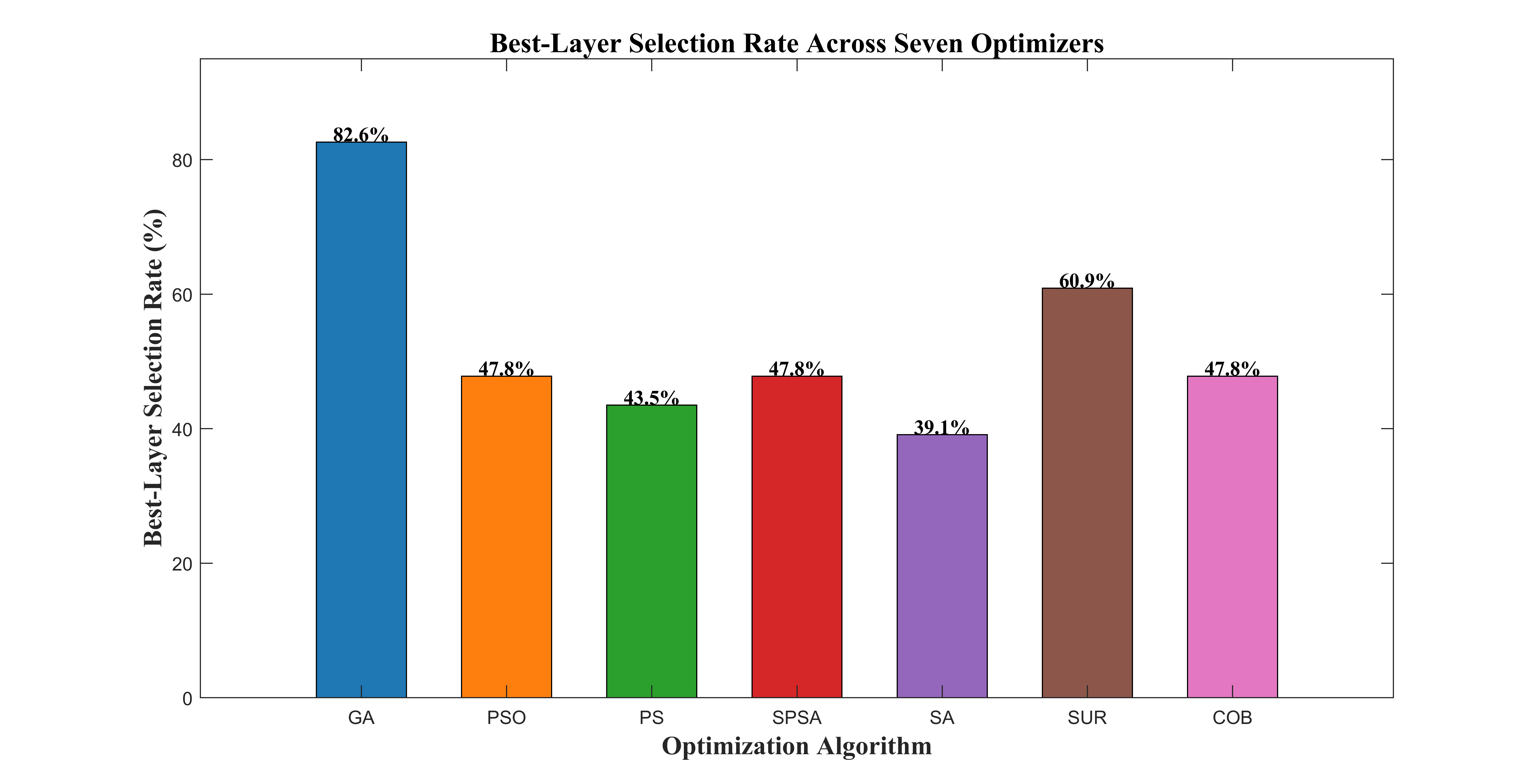}
    \caption{Best-layer selection rates across all 23 sequences, demonstrating how the optimal circuit depth varies depending on the optimizer and sequence complexity.}
    \label{fig:best_layer}
\end{figure}

Finally, the layer selection dynamics provide a fascinating insight into how algorithms interact with the ansatz depth. Figure~\ref{fig:best_layer} illustrates the best-layer selection rate. GA selects L1 most often (82.6\%) as the best layer, which initially appears contradictory to its higher mean accuracy at L2. This discrepancy arises because GA, with its strong global search, can often find the correct structure even with the limited expressivity of L1 for simpler sequences. However, the mean accuracy is pulled higher by the complex sequences that strictly require L2 to resolve deep structural correlations.

PSO, PS, and SPSA select L1, L2, and L3 in a more distributed manner, showing that different RNA sequences require varying levels of circuit depth. This distribution further reinforces the necessity of our proposed adaptive framework. In a traditional fixed-depth VQE, one would have to choose either a shallow circuit (missing complex structures) or a deep circuit (suffering from noise and barren plateaus). The Layer-CVaR-VQE framework resolves this dilemma by evaluating all depths and dynamically selecting the most suitable one for each specific RNA sequence and optimizer configuration.

In summary, the integration of population-based optimizers (GA, PSO, PS) with the adaptive Layer-CVaR-VQE architecture consistently yields the highest sensitivity, specificity, and quantum state probabilities. Gradient-based and surrogate methods remain ill-suited for the discrete, shot-noisy landscape of NISQ-era RNA folding, whereas intermediate circuit depths offer the optimal trade-off between expressivity and noise susceptibility.

\FloatBarrier
\section{Conclusion}\label{sec:conclusion}
In this study, we introduced a novel adaptive gate-based quantum framework for RNA secondary structure prediction, built upon a layered Conditional Value-at-Risk Variational Quantum Eigensolver (Layer-CVaR-VQE) architecture. By dynamically adjusting the circuit depth, our method addresses the limitations of conventional fixed-depth ansatz circuits, which often struggle to capture the varying structural complexity of different RNA sequences. We conducted an extensive benchmarking of seven classical and nature-inspired optimization algorithms, namely Genetic Algorithm (GA), Particle Swarm Optimization (PSO), Pattern Search (PS), Simulated Annealing (SA), Surrogate Optimization (SUR), Simultaneous Perturbation Stochastic Approximation (SPSA), and COBYLA, evaluating their performance against ViennaRNA, MATLAB MFE, and ProbKnot across the benchmark sequences. The empirical results demonstrate that population-based and direct-search methods significantly outperform gradient-dependent approaches in navigating the non-convex, shot-noisy energy landscape of the RNA Hamiltonian. Specifically, GA, PSO, and PS achieved the highest mean sensitivities and specificities, with GA reaching a mean sensitivity of 70.0\% and specificity of 66.0\% against ViennaRNA. Furthermore, the layer-wise analysis provided concrete evidence that intermediate circuit depths (Layer 2 and Layer 3) consistently yield superior accuracy compared to the standard single-layer CVaR-VQE, validating the hypothesis that an optimal level of ansatz expressivity is required to capture long-range structural correlations without introducing excessive noise. 

The quantum state probability analysis further reinforced these findings, revealing that GA and PSO successfully concentrate the quantum amplitude on the most thermodynamically stable bitstrings, whereas gradient-based methods like COBYLA fail to navigate the stochastic landscape. This work represents a significant step forward in quantum-assisted bioinformatics, as it provides the first comprehensive comparative study of optimization algorithms within a layered variational framework for RNA folding. It establishes a scalable foundation for algorithm-circuit co-design, demonstrating that tailored optimization strategies and adaptive circuit depths are essential for unlocking the full potential of NISQ-era quantum devices in solving NP-hard biological problems. The results confirm that while quantum hardware is still maturing, carefully designed hybrid algorithms can already offer a competitive and robust alternative to classical thermodynamic models.

Looking ahead, several promising avenues remain to be explored. First, further research should focus on upgrading the ansatz circuit gates to develop problem-specific entanglement topologies that are inherently better suited to RNA secondary structure prediction, potentially leading to even faster convergence and higher accuracy within the Layer-CVaR-VQE framework. Second, a systematic hyperparameter tuning study is required for each classical optimizer to ensure they operate at their absolute peak performance for this specific biological problem. Third, future investigations should extend the benchmark to include longer RNA sequences and complex structural motifs containing pseudoknots, which will require more sophisticated Hamiltonian formulations. Finally, as fault-tolerant quantum hardware becomes available, it will be crucial to evaluate the performance of this adaptive layered approach on real quantum computers, assessing the impact of actual device noise and qubit connectivity on the framework's overall robustness and scalability.

\section*{Acknowledgement}
The authors would like to express their sincere gratitude to Sarah Keshavarz for her invaluable assistance in designing and creating the figures used in this manuscript.

\bibliographystyle{unsrtnat}
\bibliography{references}

\end{document}